\documentclass[aps,prb,reprint,showpacs,amsmath,amssymb,nofootinbib]{revtex4-1}
\usepackage[utf8]{inputenc}
\usepackage{amsmath}
\usepackage{amssymb}
\usepackage{graphicx}

\begin{document}


\title{Geminate Exciton Fusion Fluorescence \\  as a  Probe of Triplet Exciton Transport after Singlet Fission }

\author{Eric A. Wolf}
\author{Ivan Biaggio}
\affiliation{Department of Physics, Lehigh University, Bethlehem, PA 18015, USA}

\date{\today}

\begin{abstract} 
The geminate annihilation of  two triplet excitons created by singlet exciton fission is affected by the dimensionality of transport as determined by typically anisotropic triplet exciton mobilities in organic molecular crystals. We analyze this process using a random-walk model where the  time-dynamics of the geminate annihilation probability is determined by the average exciton hopping times along the crystallographic directions. The model is then applied to the geminate fluorescence dynamics in rubrene, where the main channel for triplet-triplet annihilation is via triplet fusion and subsequent photon emission, and we identify the transitions between transport in one, two, and three dimensions.
\end{abstract}
\maketitle

Triplet exciton pairs generated from singlet exciton fission in molecular crystals can separate by independent diffusion of the two triplet excitons, and the probability of their re-encounter will determine the rate of geminate annihilation at any given time.

In materials where the triplet exciton energy is close to half the singlet exciton energy, geminate annihilation can result in triplet exciton fusion back into a singlet exciton and subsequent photon emission. The existence of fluorescence arising from this effect was recognized early on  in anthracene crystals  \cite{Kepler63, Hall63,Singh65}. 
Under pulsed illumination, triplet exciton fusion leads to a ``delayed fluorescence'' that contrasts the ``prompt fluorescence'' resulting from the radiative decay of the initial singlet exciton population \cite{Kepler63,Hall63,Singh65}. This delayed fluorescence  has been investigated on several time scales in various materials \cite{Chabr81,Funfschilling85b, Ryasnyanskiy11,Burdett12,Piland13, Jankus13}, and the time-evolution of the triplet exciton density has also been observed using other techniques like photo-induced absorption \cite{Poletayev14}. Recent interest in the topic has  grown because of the interest of multi-exciton generation processes for photovoltaics \cite{Hanna06}.

In this work, we use photons produced through the fusion of geminate triplet excitons as a probe of triplet exciton transport 
over multiple time decades, and we show that the time-dependence of geminate fusion fluorescence can provide information on the  dimensionality of triplet exciton diffusion in the crystal lattice. While spin is also important when the triplet pair has been generated in a spin-coherent state \cite{Chabr81,Funfschilling85b,Wolf18}, in the following we focus on variations in geminate fusion probability over multiple time-decades that are determined by the probability of re-encounter in the geminate triplet-pair.

We consider 
fluorescence dynamics measurements performed in a low excitation density limit where the average distance between initially photoexcited excitons is such that only geminate triplet excitons can interact with each other. These experiments  measure the time-dependence of the re-encounter probability, instead of just the  triplet density measured in absorption measurements. They also make it possible to study transport over multiple time-decades and detect events that have an exceedingly low probability of occurring but provide important information on exciton transport---typically late reencounters between triplets that have widely separated by diffusion. The rarity of such events can be compensated by appropriately long accumulation times and the sensitivity of single-photon detection. Previous studies done in this regime were confined to amorphous materials or thin films \cite{Seki05,Piland13,Jankus13,Seki18}, where the discussion of the dimensionality of diffusion is complicated either by intrinsic disorder or by the possible presence of nanocrystals \cite{Finton19}.

In  anisotropic molecular crystals, the re-encounter probability in the triplet exciton pair  must behave differently on different time scales, depending on the dimensionality of transport  \cite{Shushin2019}. Immediately after a triplet pair is created by fission, the two excitons will initially hop in the highest-mobility direction with the smallest hopping time,  diffusing in only 1 dimension (1D) before transitioning to 2 and 3 dimensions (2D and 3D) at later times. These transitions in dimensionality change how the reencounter probability evolves and lead to observable transitions in its dynamics.

We model the effect of the dimensionality of triplet exciton diffusion on geminate-annihilation dynamics using an unbiased random walk for each exciton in the triplet pair \cite{Feron2012}. Seeking for the simplest possible approach, we allow each exciton to independently hop between next-neighbors in a primitive cubic lattice, with different average hopping times in different directions. We deliberately ignore the possibility of an initial spatially correlated triplet pair \cite{Zimmerman10,Breen17} in order to focus on the essential features of later time effects.

By using the distance between the two triplet excitons as the relevant coordinate, a random walk of two excitons can be reduced to a single-particle random walk that starts at the origin, with an average hopping time that is half that of each individual exciton.
Here we modify this well-known system  \cite{Pfluegl98, Redner01} by introducing a finite average probability $\gamma$ ($0\leq \gamma \leq 1$) that a re-encounter event leads to annihilation. In this case, the probability that a re-encounter happens at time $t$ is determined by 
\begin{eqnarray}
    p_0^i(t) &=& \exp(-t/\tau_i)I_0(t/\tau_i) , \label{pZeroBessel}\\
    p_0(t) &=& p_0^{1}(t) p_0^{2}(t) p_0^{3}(t) , \label{pZero} \\
    p_\gamma(t) &=& p_\gamma(n = t/\tau) \label{pGammat}\\
    p_\gamma(n) &=& f(n) + (1-\gamma) \sum_{i=1}^{n-1} f(i) \ p_\gamma (n-i) . \label{pGamma}
\end{eqnarray}
Here, $I_0$ is the modified Bessel function of the first kind of order zero \cite{Redner01}, and $ p_{\gamma=0}^i(t)$ is the continuous time probability corresponding to  an ensemble average of all possible one-dimensional random walks \cite{Redner01}. The $p_{0}^i(t)$ are the independent probabilities that a random walk along the coordinate characterized by the hopping time $\tau_i$ is back at the origin at time $t$, and $p_0(t)$ is the combined probability that the random walk is found at the origin along each coordinate simultaneously. Finally, the re-encounter probability for finite $\gamma$, $p_\gamma(t)$, is obtained by discretizing time with the average hopping time $\tau^{-1} = \tau_1^{-1} + \tau_2^{-1}  + \tau_3^{-1}$ (we discuss alternative choices below) and using the recursive relationship (\ref{pGamma}) between the discrete-time probability $p_\gamma(n)$ that a re-encounter happens after exactly $n$ time-steps in the presence of annihilation, and the  probability $f(n)=p_1(n)$ that the \emph{first} re-encounter happens after $n$ steps. This expression is well-known in the $\gamma \to 0$ limit \cite{Pfluegl98, Redner01}.

It is useful to note that $p_0^i(t) \propto t^{-1/2}$ in the limit $t \gg \tau_i$. \cite{Redner01} It follows that for $\gamma=0$ and  a $d$-dimensional  random walk,  $p_0(t) \propto t^{-d/2}$ for large $t$, a stronger dependence on the dimensionality  than for  $\gamma=1$, where   $p_1(t) = f(t)\propto   n^{-3/2}$ for  $d=1,3$   and $p_1(t)=f(t) \propto t^{-1} \ln(t)^{-2}$  for $d=2$ (See Refs.~\onlinecite{Pfluegl98,Redner01}).

\begin{figure}[t]
\includegraphics[width=8.5cm]{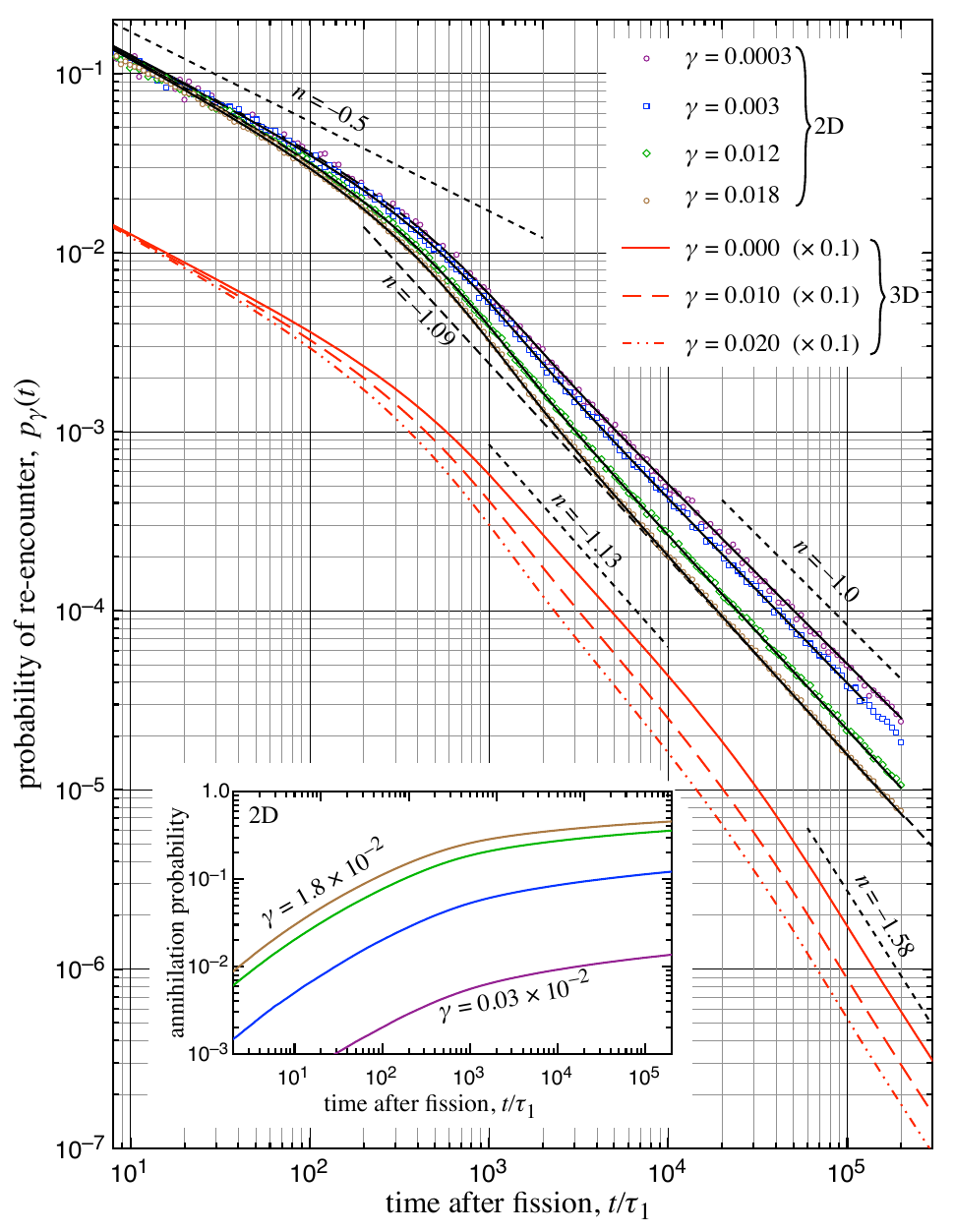} 
\caption{\label{gammadependence} Geminate re-encounter probability  after fission, for 2D and 3D transport, for different values of the average annihilation probability per re-encounter, $\gamma$. The solid curves are obtained from Eq.~\ref{pGamma}, while the dots are the result of  Monte Carlo simulations. Hopping time ratios are  $\tau_2/\tau_1 =1000$, and $\tau_3/\tau_2 = 60$ for 3D simulations. Power law regimes in the data are highlighted with displaced dashed lines labeled with the  corresponding power-law exponents. The dashed line labeled ``$n=-1.09$'' is an extrapolation from the late time behavior of the 2D  $\gamma=0.018$ simulation. The inset gives the behavior, for the 2D case,  of  the total  probability that geminate annihilation occurs before a given time. }

\end{figure}

The importance of Eq.~(\ref{pGamma}) here  is that it can be used to analytically calculate $p_\gamma(t)$ from the $p_0(t)$ in Eq.~\ref{pZero}. This is done in two steps, by  first solving for  $ f(n)=p_1(n)$ from $p_0(n)$, and then solving  for $p_\gamma(n)$ from $ f(n)$. This calculation can in general be done as described above by discretizing time  using the   average hopping time $\tau$, but we note that  practical calculations can also be done by using a longer time-step  $\Delta t$ in place of $\tau$ while at the same time using  a renormalized $\gamma \Delta t / \tau \ll 1$ in place of $\gamma$. We show in Fig.~\ref{gammadependence}  that there is an exact correspondence between this analytical method and  a Monte Carlo simulation of the random walk of the two excitons.

The Monte-Carlo simulations in Fig.~\ref{gammadependence} were done starting with two triplet excitons adjacent to each other, implementing their two independent random walks, each with average hopping times  $2 \tau_i$, and allowing  for an average annihilation probability  $\gamma$ upon a re-encounter.
We repeated the simulation about one million times, storing the time when each annihilation event occurred into logarithmically distributed time-bins to obtain the Monte Carlo data in the figure.

In this work we choose to focus on the case where triplet-triplet annihilation is relevant, but where the way in which triplet excitons can separate still leads to a meaningful singlet exciton fission probability. Hence, we do not consider the case of exciton diffusion  remaining indefinitely confined to one dimension.  In such a 1D random walk with $n$ steps, the  average number of re-encounters  is $\sim \sqrt{n}$, and geminate annihilation  is almost certain after $n \sim 1/\gamma^2$, also leading to a deviation from the initial $t^{-1/2}$ dependence, towards a steeper $t^{-3/2}$ power-law. This situation does not occur in our analysis because we  allow for a finite probability for the exciton to move out of an initial 1D behavior.

Fig.~\ref{gammadependence} highlights several features of the transition between the expected power law behaviors for 1D and 2D transport as the slower hopping time $\tau_2$ comes into play, determining an average number of steps during the initial 1D random walk of  $n=\tau_2/\tau_1$.  In the small $\gamma$ limit (here we always have $\gamma < 1/\sqrt{n}$), the asymptotic behaviors $t^{-0.5}$ and $t^{-1}$ for 1D and 2D diffusion become clearly visible  for $t \lesssim \tau_2/100$ and $t \gtrsim 10 \tau_2$, far enough away from the transition. This can be understood from the functional form of $p_0^i(x) =  \exp(-x)I_0(x)$ as defined in Eq.~\ref{pZeroBessel}: between $x \approx 0.25$ and $x=10$ it overshoots the extrapolation from its  $x \to \infty$ limit of $x^{-0.5}$, with a tangent on the log-log plot that reaches a power law with an exponent up to $n=-0.6$. A consequence of this is the fact that the  1D$\to$2D transition is always characterized by $p_\gamma(t)$ transiently becoming slightly larger than the power-law extrapolation from the 2D transport region, leading to an inflection point that becomes more prominent as $\gamma$ increases (see the dashed line extrapolation of the $\gamma=0.018$ data in  Fig.~\ref{gammadependence}). 

This feature is associated with the  constant probability per unit time of $1/\tau_2$ for an exciton to ``hop out'' of the 1D random walk, and it is important because it could lead to the appearance of an exponential decay when geminate fluorescence is measured on a limited time scale of the order of $t \sim \tau_2$ \cite{Wolf18}. 

Fig.~\ref{gammadependence} also shows the time dependence of the re-encounter probability when a 3D transport regime can be reached. There are clearly identifiable transitions as the dimensionality of the  triplet exciton transport changes with time from  1D, to 2D, to 3D. Each dimensionality is characterized by equally clear power-law regimes, with exponents slightly larger  than  predicted in the $t \to \infty$ limit of each transport regime. This effect  is stronger for the intermediate 2D transport regime (where in our example we get a power-law exponent of  $n=-1.13$ for $\gamma=0$), due both to the proximity of the transition to different dimensionalities of transport and to a larger annihilation probability.

From the re-encounter probability $p_\gamma(t)$ one  obtains a time-dependent geminate annihilation probability density $\gamma p_\gamma(t)/\tau$. The $\gamma$-dependent integral of this quantity then gives the overall probability that geminate annihilation occurs before a given time, shown in the inset of  Fig.~\ref{gammadependence}   for the case when   transport dimensionality changes from 1D to 2D. This overall geminate annihilation probability is an important parameter for applications where exciton fission yield must be optimized.

 Since the  average number of re-encounters in a 1D random walk of $n$ steps is $\sim \sqrt{n}$, much larger than for 2D random walks, the overall probability of geminate annihilation is $\sim \gamma \sqrt{\tau_2/\tau_1}$ before the 1D$\to$2D transition in the small $\gamma$ limit. This is equal to $0.1$ for $\gamma=0.3 \times 10^{-2}$  and the $\tau_2/\tau_1 = 1000$  used in Fig.~\ref{gammadependence}, in good agreement with the time-dependent geminate annihilation probability curves in the inset of the same figure. 
We note that if the 1D regime persists long enough,  then even a relatively small value of the average annihilation probability per re-encounter, $\gamma$, will lead to a significant reduction in fission efficiency -- this is shown in the inset of  Fig.~\ref{gammadependence}, where fission efficiency decreases below $50$\% for an annihilation probability per re-encounter as small as $~2$\%. It is likely that some measure of 1D transport will always be necessary to facilitate initial separation of the triplet excitons, but one must  be aware that it is also important to minimize $\gamma \sqrt{\tau_2/\tau_1}$ in order to optimize fission efficiency.

We experimentally investigated geminate fluorescence dynamics in rubrene single crystals, which have a large singlet fission efficiency, a triplet-triplet annihilation  dominated by fusion and photon emission,  large triplet lifetimes, and anisotropic triplet transport  \cite{Ryasnyanskiy11,Biaggio13b,Irkhin11}. We used single crystals grown from 99\% pure ACROS Organics rubrene powder via physical vapor transport in argon gas. We label the crystallographic axes of these crystals as $a$, $b$, and $c$ following the convention of Ref. \onlinecite{Irkhin12}. We investigated crystals as large as a few mm along the $a$ and $b$ axes, with thicknesses of the order of 100 $\mu$m along the $c$ axis. 

The samples were exposed to  150~fs pulses of 513~nm, $b$-polarized light from a Light Conversion PHAROS laser operating at $5$ kHz; this gave a time-interval between successive pulses twice as long as  the triplet exciton lifetime \cite{Ryasnyanskiy11} in rubrene, ensuring a sufficient decay of the triplet  population between illumination pulses. The beam diameter in the sample was a few mm, delivering excitation densities below $\sim 10^{19}$~m$^{-3}$, which guaranteed a fluence-independent fluorescence dynamics up to $\sim 5$~$\mu$s after photoexcitation.
The resulting geminate fluorescence dynamics was measured by time-correlated single photon counting   \cite{OConnor12}, with after-pulsing effects avoided by detecting only a maximum of one photon per cycle.
Typical integration times were of the order of 2-10 hours. 

\begin{figure}
    \centering
  \includegraphics[width = 8.5 cm]{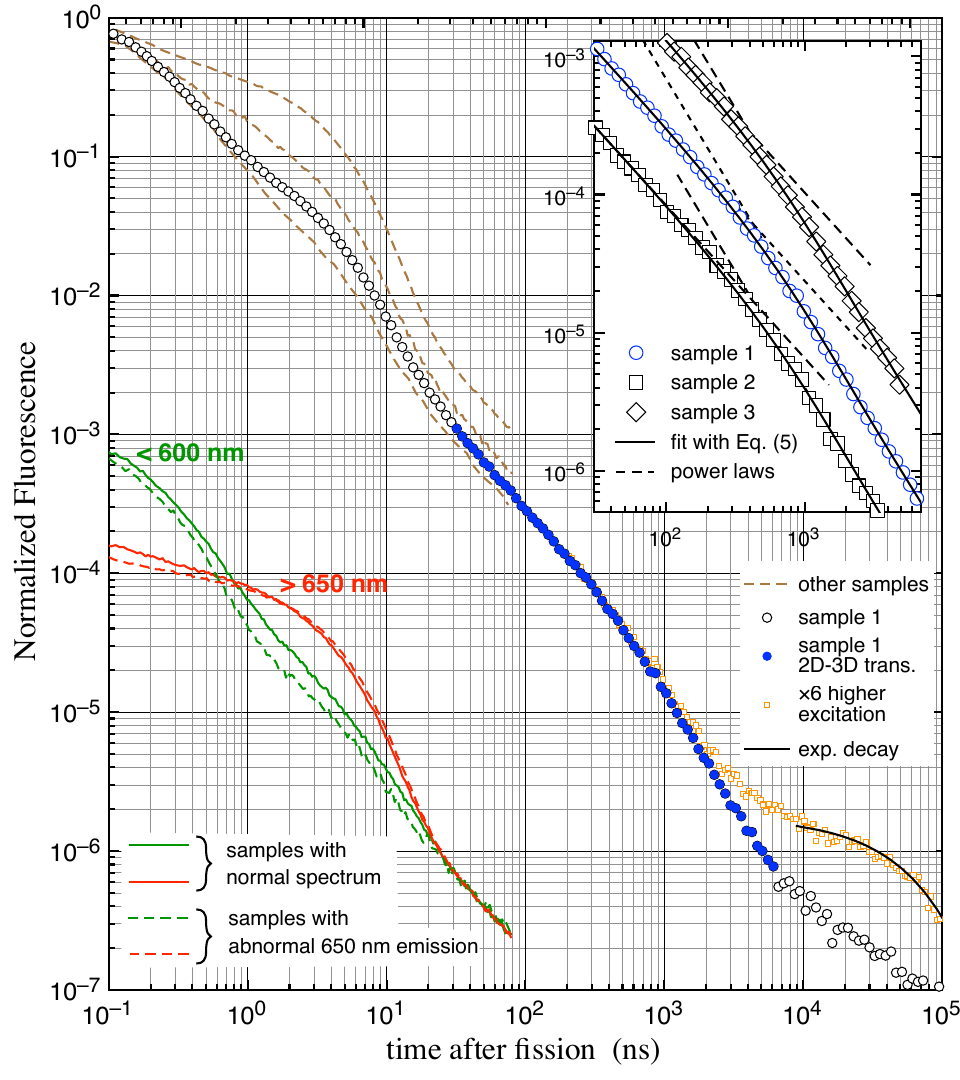} 
    \caption{Normalized multiscale geminate fluorescence dynamics in rubrene crystals  at excitation densities of $\sim 2\times 10^{19} \rm{m}^{-3}$ (round data points) and  $\sim 12 \times 10^{19} \rm{m}^{-3}$ (square, orange data points).  The higher density data can be fitted by the an exponential decay with a time constant of  60 $\mu$s (solid curve). The  filled, blue data points highlight the region where we did not see any variability from sample to sample, including those with anomalous fluorescence spectra. Sample-dependent variations at times below 100 ns (dashed lines) correlated with the presence of anomalous fluorescence emission band near 650 nm.\\ 
    Fluorescence dynamics due only to photons with wavelength less than 600 and more than 650 nm is shown by the solid (samples with normal fluorescence spectra) and dashed (samples with abnormal fluorescence spectra) curves displaced down by three decades, and normalized to the sample-independent fluorescence signal at 100 ns. \\ 
    The inset shows fluorescence dynamics from three different rubrene crystals (vertically offset for clarity) in the time window corresponding to the solid blue dots, where a transition from 2D to 3D transport takes place. Solid curves in the inset are all fits with Eq.~(\ref{FitEquation}) for $\tau_3=1.3$ $\mu s$ and $p = - 0.07$;  the dashed lines are power law extrapolations. \label{ExperimentalTransition}}
\end{figure}


Fig.~\ref{ExperimentalTransition} presents the  geminate fluorescence dynamics we obtained in several rubrene single crystals. The data was very well reproducible from sample to sample, except for the time window below 50 ns, where we observed variations in the data that correlated to the presence of an anomalous fluorescence spectrum characterized by a 650 nm emission band \cite{Mastrogiovanni14, Mitrofanov06, Mitrofanov07, Zeng07,Irkhin12}. 

In this context, we note that our apparatus also captures the picosecond-scale ``prompt'' fluorescence that is due to the radiative decay of the initially photoexcited singlet states. Even though  only a few percent at most of the photoexcited singlet states decay radiatively instead of undergoing fission \cite{Ryasnyanskiy11,Biaggio13b}, and the corresponding signal is  broadened by our limited response time, this transient signal is large enough to produce an initial decay of about 1 order of magnitude in signal intensity. 

This analysis is also supported by the data  due only to the long-wavelength fluorescence above 650 nm, which does not show the strong enhancement of the signal that is otherwise seen below 1~ns.  In fact,  the long-wavelength fluorescence is exclusively due to interaction of the triplet excitons created by fission with extraneous defects \cite{Biaggio13b, Chen11}, without any contribution from the initially photoexcited singlet states. We also note that the initial dynamics of this triplet-only fluorescence  could in principle be analyzed in terms of a 1D random walk and the expected power-law dependence, corresponding to a straight line with a slope of $-0.5$ in the log-log plot of  Fig.~\ref{ExperimentalTransition}. We cannot do so quantitatively here, however, because  our limited time-resolution makes any such slope in the first decade of Fig. 2 too difficult to interpret. In any case, we have also shown that limiting the detection to fluorescence wavelengths below 600 nm essentially eliminated any sample-dependent effects on the fluorescence dynamics, even from crystals with strong abnormal fluorescence.

 In the following we concentrate on the fluorescence dynamics that does not show any sample-to-sample variation, and in particular on the data between $\sim 100$ ns  and $\sim 10$ $\mu$s after excitation. We reach this intrinsic limit by using photoexcitation densities of the order  of $2\times10^{19}~\rm{m}^{-3} $. At higher excitation densities, one observes deviations in the fluorescence dynamics that are consistent with the onset of non-geminate annihilation. In Fig.~\ref{ExperimentalTransition} we show the deviation obtained for an excitation density of $12\times10^{19} \rm{m}^{-3}$,  with its typical exponential decay that corresponds to half the triplet exciton lifetime  \cite{Ryasnyanskiy11}. 

The most important  feature of the experimental results in Fig.~\ref{ExperimentalTransition} is that the data in the time window from $30$ ns and $5$ $\mu$s---highlighted by full data points---clearly shows the expected transition between power law regimes that is predicted by Fig.~\ref{gammadependence}. Power-law regimes at the beginning and end of this time window have exponents of  $-1.18 \pm 0.02$ and $-1.66 \pm 0.03$, respectively. In addition to this, the whole dynamics in this time window and for multiple samples can be  extremely well fitted by a slight modification of Eq.~(\ref{pZero}):
\begin{equation}
    PL(t) \propto t^{-1 + p} e^{-t/\tau} I_0(t/\tau)
    \label{FitEquation}
\end{equation}
This corresponds to Equation \ref{pZero} in the $t \gg \tau_1, \tau_2$ limit, multiplied by a factor $t^{p}$ which accounts for a possible systematic increase in the power-law exponent similar to that observed in Fig.~\ref{gammadependence}.

The inset in the top-right corner of Fig.~\ref{ExperimentalTransition} shows a simultaneous fit of the data from three different samples using Eq.~(\ref{FitEquation}), which delivers $p=-0.07$ and $\tau_3 = 1.3$ $\mu$s. The latter corresponds to an average hopping time of a triplet exciton along the least probable hopping direction (likely the crystallographic $c$-axis in rubrene \cite{da2005,Yost12}) of $2 \tau_3 = 2.6$ $\mu$s. 

Individual fits to the dynamics in multiple samples show  minimal variations of  $p$ in the range $0.05$-$0.15$ and  $\tau_3$ in the range $1{\rm -}3$ $\mu$s.

In addition to this experimental determination of the slowest hopping time $\tau_3$, we can also say something about the two faster hopping times. The $\tau_3$ that we found experimentally  is  six orders of magnitude greater than the average hopping time along the high-mobility $b$-direction, which can be estimated to be $\tau_1 \sim 3$ ps from the diffusion length of $4$ $\mu$m \cite{Irkhin11} and the triplet lifetime of $100$~$\mu$s \cite{Ryasnyanskiy11}. Then,  the fact that the power-law corresponding to 2D transport is already well-established 20 ns after the beginning of the random walk, coupled with an earlier observation of a quasi-exponential decay in the fluorescence intensity with an apparent  time-constant of 4 ns \cite{Wolf18}, indicates that the average hopping time $\tau_2$ that determines the transition from 1D to 2D transport  is very likely of the order of a few nanoseconds.

From this we see that geminate fluorescence dynamics is  sensitive to the reencounter probability of triplet excitons. This will make it necessary to consider the reencounter probability in any long-timescale investigation of delayed fluorescence in rubrene or analogous singlet fission materials. But most importantly,  geminate fusion dynamics can deliver valuable information on triplet exciton transport and the crystal lattice where it happens, all thanks to the ability to detect single photons that originate from annihilation events.\cite{bossanyi2021} In future work, this can be exploited to  understand more complicated systems, like those in which excitons interact with defects \cite{Irkhin16}.

In conclusion, we have shown that the photons emitted during geminate fusion can be effectively used as a probe of triplet exciton transport. We have also demonstrated how data obtained in this way can be  analyzed via a simple random-walk model that can be used to  extract information on average hopping times and the dimensionality of transport. While we demonstrated these ideas in rubrene, the sensitivity of single-photon counting means that the model and approach are applicable to any singlet-fission material in which geminate triplet-triplet annihilation is accompanied by some probability of photon emission. 

This opens the door to the general use of geminate-fusion fluorescence as an investigative tool  to, $e.g.$, disentangle the effect of variations in the fusion probability $\gamma$ and of variations in the triplet transport that determines   the probability of re-encounter $p_\gamma(t)$  in temperature-dependent or magnetic-field-dependent studies.

Note added in proof:  after this work was submitted, we became aware of the related work in Ref.~\onlinecite{Seki21}, which uses a completely different approach to also discuss the influence of the dimensionality of triplet transport on the geminate fluorescence.

\newpage

\begin{acknowledgments}
Acknowledgement: This research was partially supported by the U.S. Department of Energy, Office of Basic Energy Sciences, Division of Materials Sciences and Engineering under Award \# DE-SC0020981.
\end{acknowledgments}

\bibliography{RubreneEtAl}

\begin{thebibliography}{35}%
\makeatletter
\providecommand \@ifxundefined [1]{%
 \@ifx{#1\undefined}
}%
\providecommand \@ifnum [1]{%
 \ifnum #1\expandafter \@firstoftwo
 \else \expandafter \@secondoftwo
 \fi
}%
\providecommand \@ifx [1]{%
 \ifx #1\expandafter \@firstoftwo
 \else \expandafter \@secondoftwo
 \fi
}%
\providecommand \natexlab [1]{#1}%
\providecommand \enquote  [1]{``#1''}%
\providecommand \bibnamefont  [1]{#1}%
\providecommand \bibfnamefont [1]{#1}%
\providecommand \citenamefont [1]{#1}%
\providecommand \href@noop [0]{\@secondoftwo}%
\providecommand \href [0]{\begingroup \@sanitize@url \@href}%
\providecommand \@href[1]{\@@startlink{#1}\@@href}%
\providecommand \@@href[1]{\endgroup#1\@@endlink}%
\providecommand \@sanitize@url [0]{\catcode `\\12\catcode `\$12\catcode
  `\&12\catcode `\#12\catcode `\^12\catcode `\_12\catcode `\%12\relax}%
\providecommand \@@startlink[1]{}%
\providecommand \@@endlink[0]{}%
\providecommand \url  [0]{\begingroup\@sanitize@url \@url }%
\providecommand \@url [1]{\endgroup\@href {#1}{\urlprefix }}%
\providecommand \urlprefix  [0]{URL }%
\providecommand \Eprint [0]{\href }%
\providecommand \doibase [0]{http://dx.doi.org/}%
\providecommand \selectlanguage [0]{\@gobble}%
\providecommand \bibinfo  [0]{\@secondoftwo}%
\providecommand \bibfield  [0]{\@secondoftwo}%
\providecommand \translation [1]{[#1]}%
\providecommand \BibitemOpen [0]{}%
\providecommand \bibitemStop [0]{}%
\providecommand \bibitemNoStop [0]{.\EOS\space}%
\providecommand \EOS [0]{\spacefactor3000\relax}%
\providecommand \BibitemShut  [1]{\csname bibitem#1\endcsname}%
\let\auto@bib@innerbib\@empty
\bibitem [{\citenamefont {Kepler}\ \emph {et~al.}(1963)\citenamefont {Kepler},
  \citenamefont {Caris}, \citenamefont {Avakian},\ and\ \citenamefont
  {Abramson}}]{Kepler63}%
  \BibitemOpen
  \bibfield  {author} {\bibinfo {author} {\bibfnamefont {R.~G.}\ \bibnamefont
  {Kepler}}, \bibinfo {author} {\bibfnamefont {J.~C.}\ \bibnamefont {Caris}},
  \bibinfo {author} {\bibfnamefont {P.}~\bibnamefont {Avakian}}, \ and\
  \bibinfo {author} {\bibfnamefont {E.}~\bibnamefont {Abramson}},\ }\href@noop
  {} {\bibfield  {journal} {\bibinfo  {journal} {Phys. Rev. Lett.}\ }\textbf
  {\bibinfo {volume} {10}},\ \bibinfo {pages} {400} (\bibinfo {year}
  {1963})}\BibitemShut {NoStop}%
\bibitem [{\citenamefont {Hall}\ \emph {et~al.}(1963)\citenamefont {Hall},
  \citenamefont {D.A.Jennings},\ and\ \citenamefont {R.M.McClintock}}]{Hall63}%
  \BibitemOpen
  \bibfield  {author} {\bibinfo {author} {\bibfnamefont {J.}~\bibnamefont
  {Hall}}, \bibinfo {author} {\bibnamefont {D.A.Jennings}}, \ and\ \bibinfo
  {author} {\bibnamefont {R.M.McClintock}},\ }\href@noop {} {\bibfield
  {journal} {\bibinfo  {journal} {Phys. Rev. Lett.}\ }\textbf {\bibinfo
  {volume} {11}},\ \bibinfo {pages} {364} (\bibinfo {year} {1963})}\BibitemShut
  {NoStop}%
\bibitem [{\citenamefont {Singh}\ \emph {et~al.}(1965)\citenamefont {Singh},
  \citenamefont {Jones}, \citenamefont {Siebrand}, \citenamefont {Stoicheff},\
  and\ \citenamefont {Schneider}}]{Singh65}%
  \BibitemOpen
  \bibfield  {author} {\bibinfo {author} {\bibfnamefont {S.}~\bibnamefont
  {Singh}}, \bibinfo {author} {\bibfnamefont {W.}~\bibnamefont {Jones}},
  \bibinfo {author} {\bibfnamefont {W.}~\bibnamefont {Siebrand}}, \bibinfo
  {author} {\bibfnamefont {B.}~\bibnamefont {Stoicheff}}, \ and\ \bibinfo
  {author} {\bibfnamefont {W.}~\bibnamefont {Schneider}},\ }\href@noop {}
  {\bibfield  {journal} {\bibinfo  {journal} {The Journal of Chemical Physics}\
  }\textbf {\bibinfo {volume} {42}},\ \bibinfo {pages} {330} (\bibinfo {year}
  {1965})}\BibitemShut {NoStop}%
\bibitem [{\citenamefont {Chabr}\ \emph {et~al.}(1981)\citenamefont {Chabr},
  \citenamefont {Wild}, \citenamefont {Funfschilling},\ and\ \citenamefont
  {Zschokke-Granacher}}]{Chabr81}%
  \BibitemOpen
  \bibfield  {author} {\bibinfo {author} {\bibfnamefont {M.}~\bibnamefont
  {Chabr}}, \bibinfo {author} {\bibfnamefont {U.}~\bibnamefont {Wild}},
  \bibinfo {author} {\bibfnamefont {J.}~\bibnamefont {Funfschilling}}, \ and\
  \bibinfo {author} {\bibfnamefont {I.}~\bibnamefont {Zschokke-Granacher}},\
  }\href@noop {} {\bibfield  {journal} {\bibinfo  {journal} {Chemical Physics}\
  }\textbf {\bibinfo {volume} {57}},\ \bibinfo {pages} {425} (\bibinfo {year}
  {1981})}\BibitemShut {NoStop}%
\bibitem [{\citenamefont {Funfschilling}\ \emph {et~al.}(1985)\citenamefont
  {Funfschilling}, \citenamefont {Zschokkegranacher}, \citenamefont
  {Canonica},\ and\ \citenamefont {Wild}}]{Funfschilling85b}%
  \BibitemOpen
  \bibfield  {author} {\bibinfo {author} {\bibfnamefont {J.}~\bibnamefont
  {Funfschilling}}, \bibinfo {author} {\bibfnamefont {I.}~\bibnamefont
  {Zschokkegranacher}}, \bibinfo {author} {\bibfnamefont {S.}~\bibnamefont
  {Canonica}}, \ and\ \bibinfo {author} {\bibfnamefont {U.~P.}\ \bibnamefont
  {Wild}},\ }\href@noop {} {\bibfield  {journal} {\bibinfo  {journal}
  {Helvetica Physica Acta}\ }\textbf {\bibinfo {volume} {58}},\ \bibinfo
  {pages} {347} (\bibinfo {year} {1985})}\BibitemShut {NoStop}%
\bibitem [{\citenamefont {Ryasnyanskiy}\ and\ \citenamefont
  {Biaggio}(2011)}]{Ryasnyanskiy11}%
  \BibitemOpen
  \bibfield  {author} {\bibinfo {author} {\bibfnamefont {A.}~\bibnamefont
  {Ryasnyanskiy}}\ and\ \bibinfo {author} {\bibfnamefont {I.}~\bibnamefont
  {Biaggio}},\ }\href {\doibase 10.1103/PhysRevB.84.193203} {\bibfield
  {journal} {\bibinfo  {journal} {Phys. Rev. B}\ }\textbf {\bibinfo {volume}
  {84}},\ \bibinfo {pages} {193203} (\bibinfo {year} {2011})}\BibitemShut
  {NoStop}%
\bibitem [{\citenamefont {Burdett}\ and\ \citenamefont
  {Bardeen}(2012)}]{Burdett12}%
  \BibitemOpen
  \bibfield  {author} {\bibinfo {author} {\bibfnamefont {J.~J.}\ \bibnamefont
  {Burdett}}\ and\ \bibinfo {author} {\bibfnamefont {C.~J.}\ \bibnamefont
  {Bardeen}},\ }\href {\doibase 10.1021/ja301683w} {\bibfield  {journal}
  {\bibinfo  {journal} {Journal of the American Chemical Society}\ }\textbf
  {\bibinfo {volume} {134}},\ \bibinfo {pages} {8597} (\bibinfo {year}
  {2012})}\BibitemShut {NoStop}%
\bibitem [{\citenamefont {Piland}\ \emph {et~al.}(2013)\citenamefont {Piland},
  \citenamefont {Burdett}, \citenamefont {Kurunthu},\ and\ \citenamefont
  {Bardeen}}]{Piland13}%
  \BibitemOpen
  \bibfield  {author} {\bibinfo {author} {\bibfnamefont {G.~B.}\ \bibnamefont
  {Piland}}, \bibinfo {author} {\bibfnamefont {J.~J.}\ \bibnamefont {Burdett}},
  \bibinfo {author} {\bibfnamefont {D.}~\bibnamefont {Kurunthu}}, \ and\
  \bibinfo {author} {\bibfnamefont {C.~J.}\ \bibnamefont {Bardeen}},\ }\href
  {\doibase 10.1021/jp309286v} {\bibfield  {journal} {\bibinfo  {journal} {The
  Journal of Physical Chemistry C}\ }\textbf {\bibinfo {volume} {117}},\
  \bibinfo {pages} {1224} (\bibinfo {year} {2013})}\BibitemShut {NoStop}%
\bibitem [{\citenamefont {Jankus}\ \emph {et~al.}(2013)\citenamefont {Jankus},
  \citenamefont {Snedden}, \citenamefont {Bright}, \citenamefont {Arac},
  \citenamefont {Dai},\ and\ \citenamefont {Monkman}}]{Jankus13}%
  \BibitemOpen
  \bibfield  {author} {\bibinfo {author} {\bibfnamefont {V.}~\bibnamefont
  {Jankus}}, \bibinfo {author} {\bibfnamefont {E.~W.}\ \bibnamefont {Snedden}},
  \bibinfo {author} {\bibfnamefont {D.~W.}\ \bibnamefont {Bright}}, \bibinfo
  {author} {\bibfnamefont {E.}~\bibnamefont {Arac}}, \bibinfo {author}
  {\bibfnamefont {D.}~\bibnamefont {Dai}}, \ and\ \bibinfo {author}
  {\bibfnamefont {A.~P.}\ \bibnamefont {Monkman}},\ }\href@noop {} {\bibfield
  {journal} {\bibinfo  {journal} {Phys. Rev. B}\ }\textbf {\bibinfo {volume}
  {87}},\ \bibinfo {pages} {224202} (\bibinfo {year} {2013})}\BibitemShut
  {NoStop}%
\bibitem [{\citenamefont {Poletayev}\ \emph {et~al.}(2014)\citenamefont
  {Poletayev}, \citenamefont {Clark}, \citenamefont {Wilson}, \citenamefont
  {Rao}, \citenamefont {Makino}, \citenamefont {Hotta},\ and\ \citenamefont
  {Friend}}]{Poletayev14}%
  \BibitemOpen
  \bibfield  {author} {\bibinfo {author} {\bibfnamefont {A.~D.}\ \bibnamefont
  {Poletayev}}, \bibinfo {author} {\bibfnamefont {J.}~\bibnamefont {Clark}},
  \bibinfo {author} {\bibfnamefont {M.~W.~B.}\ \bibnamefont {Wilson}}, \bibinfo
  {author} {\bibfnamefont {A.}~\bibnamefont {Rao}}, \bibinfo {author}
  {\bibfnamefont {Y.}~\bibnamefont {Makino}}, \bibinfo {author} {\bibfnamefont
  {S.}~\bibnamefont {Hotta}}, \ and\ \bibinfo {author} {\bibfnamefont {R.~H.}\
  \bibnamefont {Friend}},\ }\bibfield  {booktitle} {\emph {\bibinfo {booktitle}
  {Advanced Materials}},\ }\href {\doibase 10.1002/adma.201302427} {\bibfield
  {journal} {\bibinfo  {journal} {Advanced Materials}\ }\textbf {\bibinfo
  {volume} {26}},\ \bibinfo {pages} {919} (\bibinfo {year} {2014})}\BibitemShut
  {NoStop}%
\bibitem [{\citenamefont {Hanna}\ and\ \citenamefont {Nozik}(2006)}]{Hanna06}%
  \BibitemOpen
  \bibfield  {author} {\bibinfo {author} {\bibfnamefont {M.~C.}\ \bibnamefont
  {Hanna}}\ and\ \bibinfo {author} {\bibfnamefont {A.~J.}\ \bibnamefont
  {Nozik}},\ }\bibfield  {booktitle} {\emph {\bibinfo {booktitle} {Journal of
  Applied Physics}},\ }\href {\doibase 10.1063/1.2356795} {\bibfield  {journal}
  {\bibinfo  {journal} {Journal of Applied Physics}\ }\textbf {\bibinfo
  {volume} {100}},\ \bibinfo {pages} {074510} (\bibinfo {year}
  {2006})}\BibitemShut {NoStop}%
\bibitem [{\citenamefont {Wolf}\ \emph {et~al.}(2018)\citenamefont {Wolf},
  \citenamefont {Finton}, \citenamefont {Zoutenbier},\ and\ \citenamefont
  {Biaggio}}]{Wolf18}%
  \BibitemOpen
  \bibfield  {author} {\bibinfo {author} {\bibfnamefont {E.~A.}\ \bibnamefont
  {Wolf}}, \bibinfo {author} {\bibfnamefont {D.~M.}\ \bibnamefont {Finton}},
  \bibinfo {author} {\bibfnamefont {V.}~\bibnamefont {Zoutenbier}}, \ and\
  \bibinfo {author} {\bibfnamefont {I.}~\bibnamefont {Biaggio}},\ }\href
  {\doibase 10.1063/1.5020652} {\bibfield  {journal} {\bibinfo  {journal}
  {Applied Physics Letters}\ }\textbf {\bibinfo {volume} {112}},\ \bibinfo
  {pages} {083301} (\bibinfo {year} {2018})}\BibitemShut {NoStop}%
\bibitem [{\citenamefont {Seki}\ \emph {et~al.}(2005)\citenamefont {Seki},
  \citenamefont {Murayama},\ and\ \citenamefont {Tachiya}}]{Seki05}%
  \BibitemOpen
  \bibfield  {author} {\bibinfo {author} {\bibfnamefont {K.}~\bibnamefont
  {Seki}}, \bibinfo {author} {\bibfnamefont {K.}~\bibnamefont {Murayama}}, \
  and\ \bibinfo {author} {\bibfnamefont {M.}~\bibnamefont {Tachiya}},\ }\href
  {\doibase 10.1103/PhysRevB.71.235212} {\bibfield  {journal} {\bibinfo
  {journal} {Phys. Rev. B}\ }\textbf {\bibinfo {volume} {71}},\ \bibinfo
  {pages} {235212} (\bibinfo {year} {2005})}\BibitemShut {NoStop}%
\bibitem [{\citenamefont {Seki}\ \emph {et~al.}(2018)\citenamefont {Seki},
  \citenamefont {Sonoda},\ and\ \citenamefont {Katoh}}]{Seki18}%
  \BibitemOpen
  \bibfield  {author} {\bibinfo {author} {\bibfnamefont {K.}~\bibnamefont
  {Seki}}, \bibinfo {author} {\bibfnamefont {Y.}~\bibnamefont {Sonoda}}, \ and\
  \bibinfo {author} {\bibfnamefont {R.}~\bibnamefont {Katoh}},\ }\href
  {\doibase 10.1021/acs.jpcc.8b02234} {\bibfield  {journal} {\bibinfo
  {journal} {The Journal of Physical Chemistry C}\ }\textbf {\bibinfo {volume}
  {122}},\ \bibinfo {pages} {11659} (\bibinfo {year} {2018})},\ \Eprint
  {http://arxiv.org/abs/https://doi.org/10.1021/acs.jpcc.8b02234}
  {https://doi.org/10.1021/acs.jpcc.8b02234} \BibitemShut {NoStop}%
\bibitem [{\citenamefont {Finton}\ \emph {et~al.}(2019)\citenamefont {Finton},
  \citenamefont {Wolf}, \citenamefont {Zoutenbier}, \citenamefont {Ward},\ and\
  \citenamefont {Biaggio}}]{Finton19}%
  \BibitemOpen
  \bibfield  {author} {\bibinfo {author} {\bibfnamefont {D.~M.}\ \bibnamefont
  {Finton}}, \bibinfo {author} {\bibfnamefont {E.~A.}\ \bibnamefont {Wolf}},
  \bibinfo {author} {\bibfnamefont {V.~S.}\ \bibnamefont {Zoutenbier}},
  \bibinfo {author} {\bibfnamefont {K.~A.}\ \bibnamefont {Ward}}, \ and\
  \bibinfo {author} {\bibfnamefont {I.}~\bibnamefont {Biaggio}},\ }\bibfield
  {booktitle} {\emph {\bibinfo {booktitle} {AIP Advances}},\ }\href {\doibase
  10.1063/1.5118942} {\bibfield  {journal} {\bibinfo  {journal} {AIP Advances}\
  }\textbf {\bibinfo {volume} {9}},\ \bibinfo {pages} {095027} (\bibinfo {year}
  {2019})}\BibitemShut {NoStop}%
\bibitem [{\citenamefont {Shushin}(2019)}]{Shushin2019}%
  \BibitemOpen
  \bibfield  {author} {\bibinfo {author} {\bibfnamefont {A.~I.}\ \bibnamefont
  {Shushin}},\ }\href {\doibase 10.1063/1.5099667} {\bibfield  {journal}
  {\bibinfo  {journal} {The Journal of Chemical Physics}\ }\textbf {\bibinfo
  {volume} {151}},\ \bibinfo {pages} {034103} (\bibinfo {year} {2019})},\
  \Eprint {http://arxiv.org/abs/https://doi.org/10.1063/1.5099667}
  {https://doi.org/10.1063/1.5099667} \BibitemShut {NoStop}%
\bibitem [{\citenamefont {Feron}\ \emph {et~al.}(2012)\citenamefont {Feron},
  \citenamefont {Zhou}, \citenamefont {Belcher},\ and\ \citenamefont
  {Dastoor}}]{Feron2012}%
  \BibitemOpen
  \bibfield  {author} {\bibinfo {author} {\bibfnamefont {K.}~\bibnamefont
  {Feron}}, \bibinfo {author} {\bibfnamefont {X.}~\bibnamefont {Zhou}},
  \bibinfo {author} {\bibfnamefont {W.~J.}\ \bibnamefont {Belcher}}, \ and\
  \bibinfo {author} {\bibfnamefont {P.~C.}\ \bibnamefont {Dastoor}},\ }\href
  {\doibase 10.1063/1.3687373} {\bibfield  {journal} {\bibinfo  {journal}
  {Journal of Applied Physics}\ }\textbf {\bibinfo {volume} {111}},\ \bibinfo
  {pages} {044510} (\bibinfo {year} {2012})},\ \Eprint
  {http://arxiv.org/abs/https://doi.org/10.1063/1.3687373}
  {https://doi.org/10.1063/1.3687373} \BibitemShut {NoStop}%
\bibitem [{\citenamefont {Zimmerman}\ \emph {et~al.}(2010)\citenamefont
  {Zimmerman}, \citenamefont {Zhang},\ and\ \citenamefont
  {Musgrave}}]{Zimmerman10}%
  \BibitemOpen
  \bibfield  {author} {\bibinfo {author} {\bibfnamefont {P.~M.}\ \bibnamefont
  {Zimmerman}}, \bibinfo {author} {\bibfnamefont {Z.}~\bibnamefont {Zhang}}, \
  and\ \bibinfo {author} {\bibfnamefont {C.~B.}\ \bibnamefont {Musgrave}},\
  }\href@noop {} {\bibfield  {journal} {\bibinfo  {journal} {Nat. Chem.}\
  }\textbf {\bibinfo {volume} {2}},\ \bibinfo {pages} {648} (\bibinfo {year}
  {2010})}\BibitemShut {NoStop}%
\bibitem [{\citenamefont {Breen}\ \emph {et~al.}(2017)\citenamefont {Breen},
  \citenamefont {Tempelaar}, \citenamefont {Bizimana}, \citenamefont {Kloss},
  \citenamefont {Reichman},\ and\ \citenamefont {Turner}}]{Breen17}%
  \BibitemOpen
  \bibfield  {author} {\bibinfo {author} {\bibfnamefont {I.}~\bibnamefont
  {Breen}}, \bibinfo {author} {\bibfnamefont {R.}~\bibnamefont {Tempelaar}},
  \bibinfo {author} {\bibfnamefont {L.~A.}\ \bibnamefont {Bizimana}}, \bibinfo
  {author} {\bibfnamefont {B.}~\bibnamefont {Kloss}}, \bibinfo {author}
  {\bibfnamefont {D.~R.}\ \bibnamefont {Reichman}}, \ and\ \bibinfo {author}
  {\bibfnamefont {D.~B.}\ \bibnamefont {Turner}},\ }\href@noop {} {\bibfield
  {journal} {\bibinfo  {journal} {Journal of the American Chemical Society}\
  }\textbf {\bibinfo {volume} {139}},\ \bibinfo {pages} {11745} (\bibinfo
  {year} {2017})}\BibitemShut {NoStop}%
\bibitem [{\citenamefont {Pfluegl}\ and\ \citenamefont
  {Silbey}(1998)}]{Pfluegl98}%
  \BibitemOpen
  \bibfield  {author} {\bibinfo {author} {\bibfnamefont {W.}~\bibnamefont
  {Pfluegl}}\ and\ \bibinfo {author} {\bibfnamefont {R.~J.}\ \bibnamefont
  {Silbey}},\ }\href {\doibase 10.1103/PhysRevE.58.4128} {\bibfield  {journal}
  {\bibinfo  {journal} {Phys. Rev. E}\ }\textbf {\bibinfo {volume} {58}},\
  \bibinfo {pages} {4128} (\bibinfo {year} {1998})}\BibitemShut {NoStop}%
\bibitem [{\citenamefont {Redner}(2001)}]{Redner01}%
  \BibitemOpen
  \bibfield  {author} {\bibinfo {author} {\bibfnamefont {S.}~\bibnamefont
  {Redner}},\ }\href {\doibase 10.1017/CBO9780511606014} {\emph {\bibinfo
  {title} {A Guide to First-Passage Processes}}}\ (\bibinfo  {publisher}
  {Cambridge University Press},\ \bibinfo {year} {2001})\BibitemShut {NoStop}%
\bibitem [{\citenamefont {Biaggio}\ and\ \citenamefont
  {Irkhin}(2013)}]{Biaggio13b}%
  \BibitemOpen
  \bibfield  {author} {\bibinfo {author} {\bibfnamefont {I.}~\bibnamefont
  {Biaggio}}\ and\ \bibinfo {author} {\bibfnamefont {P.}~\bibnamefont
  {Irkhin}},\ }\href@noop {} {\bibfield  {journal} {\bibinfo  {journal}
  {Applied Physics Letters}\ }\textbf {\bibinfo {volume} {103}},\ \bibinfo
  {pages} {263301} (\bibinfo {year} {2013})}\BibitemShut {NoStop}%
\bibitem [{\citenamefont {Irkhin}\ and\ \citenamefont
  {Biaggio}(2011)}]{Irkhin11}%
  \BibitemOpen
  \bibfield  {author} {\bibinfo {author} {\bibfnamefont {P.}~\bibnamefont
  {Irkhin}}\ and\ \bibinfo {author} {\bibfnamefont {I.}~\bibnamefont
  {Biaggio}},\ }\href {\doibase 10.1103/PhysRevLett.107.017402} {\bibfield
  {journal} {\bibinfo  {journal} {Phys. Rev. Lett.}\ }\textbf {\bibinfo
  {volume} {107}},\ \bibinfo {pages} {017402} (\bibinfo {year}
  {2011})}\BibitemShut {NoStop}%
\bibitem [{\citenamefont {Irkhin}\ \emph {et~al.}(2012)\citenamefont {Irkhin},
  \citenamefont {Ryasnyanskiy}, \citenamefont {Koehler},\ and\ \citenamefont
  {Biaggio}}]{Irkhin12}%
  \BibitemOpen
  \bibfield  {author} {\bibinfo {author} {\bibfnamefont {P.}~\bibnamefont
  {Irkhin}}, \bibinfo {author} {\bibfnamefont {A.}~\bibnamefont
  {Ryasnyanskiy}}, \bibinfo {author} {\bibfnamefont {M.}~\bibnamefont
  {Koehler}}, \ and\ \bibinfo {author} {\bibfnamefont {I.}~\bibnamefont
  {Biaggio}},\ }\href {\doibase 10.1103/PhysRevB.86.085143} {\bibfield
  {journal} {\bibinfo  {journal} {Phys. Rev. B}\ }\textbf {\bibinfo {volume}
  {86}},\ \bibinfo {pages} {085143} (\bibinfo {year} {2012})}\BibitemShut
  {NoStop}%
\bibitem [{\citenamefont {O'Connor}(2012)}]{OConnor12}%
  \BibitemOpen
  \bibfield  {author} {\bibinfo {author} {\bibfnamefont {D.}~\bibnamefont
  {O'Connor}},\ }\href@noop {} {\emph {\bibinfo {title} {Time-correlated single
  photon counting}}}\ (\bibinfo  {publisher} {Academic Press},\ \bibinfo {year}
  {2012})\BibitemShut {NoStop}%
\bibitem [{\citenamefont {Mastrogiovanni}\ \emph {et~al.}(2014)\citenamefont
  {Mastrogiovanni}, \citenamefont {Mayer}, \citenamefont {Wan}, \citenamefont
  {Vishnyakov}, \citenamefont {Neimark}, \citenamefont {Podzorov},
  \citenamefont {Feldman},\ and\ \citenamefont {Garfunkel}}]{Mastrogiovanni14}%
  \BibitemOpen
  \bibfield  {author} {\bibinfo {author} {\bibfnamefont {D.~D.~T.}\
  \bibnamefont {Mastrogiovanni}}, \bibinfo {author} {\bibfnamefont
  {J.}~\bibnamefont {Mayer}}, \bibinfo {author} {\bibfnamefont {A.~S.}\
  \bibnamefont {Wan}}, \bibinfo {author} {\bibfnamefont {A.}~\bibnamefont
  {Vishnyakov}}, \bibinfo {author} {\bibfnamefont {A.~V.}\ \bibnamefont
  {Neimark}}, \bibinfo {author} {\bibfnamefont {V.}~\bibnamefont {Podzorov}},
  \bibinfo {author} {\bibfnamefont {L.~C.}\ \bibnamefont {Feldman}}, \ and\
  \bibinfo {author} {\bibfnamefont {E.}~\bibnamefont {Garfunkel}},\ }\href
  {http://dx.doi.org/10.1038/srep04753} {\bibfield  {journal} {\bibinfo
  {journal} {Scientific Reports}\ }\textbf {\bibinfo {volume} {4}},\ \bibinfo
  {pages} {04753} (\bibinfo {year} {2014})}\BibitemShut {NoStop}%
\bibitem [{\citenamefont {Mitrofanov}\ \emph {et~al.}(2006)\citenamefont
  {Mitrofanov}, \citenamefont {Lang}, \citenamefont {Kloc}, \citenamefont
  {Wikberg}, \citenamefont {Siegrist}, \citenamefont {So}, \citenamefont
  {Sergent},\ and\ \citenamefont {Ramirez}}]{Mitrofanov06}%
  \BibitemOpen
  \bibfield  {author} {\bibinfo {author} {\bibfnamefont {O.}~\bibnamefont
  {Mitrofanov}}, \bibinfo {author} {\bibfnamefont {D.~V.}\ \bibnamefont
  {Lang}}, \bibinfo {author} {\bibfnamefont {C.}~\bibnamefont {Kloc}}, \bibinfo
  {author} {\bibfnamefont {J.~M.}\ \bibnamefont {Wikberg}}, \bibinfo {author}
  {\bibfnamefont {T.}~\bibnamefont {Siegrist}}, \bibinfo {author}
  {\bibfnamefont {W.-Y.}\ \bibnamefont {So}}, \bibinfo {author} {\bibfnamefont
  {M.~A.}\ \bibnamefont {Sergent}}, \ and\ \bibinfo {author} {\bibfnamefont
  {A.~P.}\ \bibnamefont {Ramirez}},\ }\href
  {http://link.aps.org/abstract/PRL/v97/e166601} {\bibfield  {journal}
  {\bibinfo  {journal} {Phys. Rev. Lett.}\ }\textbf {\bibinfo {volume} {97}},\
  \bibinfo {pages} {166601} (\bibinfo {year} {2006})}\BibitemShut {NoStop}%
\bibitem [{\citenamefont {Mitrofanov}\ \emph {et~al.}(2007)\citenamefont
  {Mitrofanov}, \citenamefont {Kloc}, \citenamefont {Siegrist}, \citenamefont
  {Lang}, \citenamefont {So},\ and\ \citenamefont {Ramirez}}]{Mitrofanov07}%
  \BibitemOpen
  \bibfield  {author} {\bibinfo {author} {\bibfnamefont {O.}~\bibnamefont
  {Mitrofanov}}, \bibinfo {author} {\bibfnamefont {C.}~\bibnamefont {Kloc}},
  \bibinfo {author} {\bibfnamefont {T.}~\bibnamefont {Siegrist}}, \bibinfo
  {author} {\bibfnamefont {D.~V.}\ \bibnamefont {Lang}}, \bibinfo {author}
  {\bibfnamefont {W.-Y.}\ \bibnamefont {So}}, \ and\ \bibinfo {author}
  {\bibfnamefont {A.~P.}\ \bibnamefont {Ramirez}},\ }\href {\doibase
  10.1063/1.2815939} {\bibfield  {journal} {\bibinfo  {journal} {Applied
  Physics Letters}\ }\textbf {\bibinfo {volume} {91}},\ \bibinfo {eid} {212106}
  (\bibinfo {year} {2007})}\BibitemShut {NoStop}%
\bibitem [{\citenamefont {Zeng}\ \emph {et~al.}(2007)\citenamefont {Zeng},
  \citenamefont {Zhang}, \citenamefont {Duan}, \citenamefont {Wang},
  \citenamefont {Dong},\ and\ \citenamefont {Qiu}}]{Zeng07}%
  \BibitemOpen
  \bibfield  {author} {\bibinfo {author} {\bibfnamefont {X.}~\bibnamefont
  {Zeng}}, \bibinfo {author} {\bibfnamefont {D.}~\bibnamefont {Zhang}},
  \bibinfo {author} {\bibfnamefont {L.}~\bibnamefont {Duan}}, \bibinfo {author}
  {\bibfnamefont {L.}~\bibnamefont {Wang}}, \bibinfo {author} {\bibfnamefont
  {G.}~\bibnamefont {Dong}}, \ and\ \bibinfo {author} {\bibfnamefont
  {Y.}~\bibnamefont {Qiu}},\ }\href {\doibase DOI:
  10.1016/j.apsusc.2007.01.008} {\bibfield  {journal} {\bibinfo  {journal}
  {Applied Surface Science}\ }\textbf {\bibinfo {volume} {253}},\ \bibinfo
  {pages} {6047 } (\bibinfo {year} {2007})}\BibitemShut {NoStop}%
\bibitem [{\citenamefont {Chen}\ \emph {et~al.}(2011)\citenamefont {Chen},
  \citenamefont {Lee}, \citenamefont {Fu},\ and\ \citenamefont
  {Podzorov}}]{Chen11}%
  \BibitemOpen
  \bibfield  {author} {\bibinfo {author} {\bibfnamefont {Y.}~\bibnamefont
  {Chen}}, \bibinfo {author} {\bibfnamefont {B.}~\bibnamefont {Lee}}, \bibinfo
  {author} {\bibfnamefont {D.}~\bibnamefont {Fu}}, \ and\ \bibinfo {author}
  {\bibfnamefont {V.}~\bibnamefont {Podzorov}},\ }\href {\doibase
  10.1002/adma.201102294} {\bibfield  {journal} {\bibinfo  {journal} {Advanced
  Materials}\ ,\ \bibinfo {pages} {n/a}} (\bibinfo {year} {2011})}\BibitemShut
  {NoStop}%
\bibitem [{\citenamefont {da~Silva~Filho}\ \emph {et~al.}(2005)\citenamefont
  {da~Silva~Filho}, \citenamefont {Kim},\ and\ \citenamefont
  {Br{\'e}das}}]{da2005}%
  \BibitemOpen
  \bibfield  {author} {\bibinfo {author} {\bibfnamefont {D.~A.}\ \bibnamefont
  {da~Silva~Filho}}, \bibinfo {author} {\bibfnamefont {E.-G.}\ \bibnamefont
  {Kim}}, \ and\ \bibinfo {author} {\bibfnamefont {J.-L.}\ \bibnamefont
  {Br{\'e}das}},\ }\href@noop {} {\bibfield  {journal} {\bibinfo  {journal}
  {Advanced Materials}\ }\textbf {\bibinfo {volume} {17}},\ \bibinfo {pages}
  {1072} (\bibinfo {year} {2005})}\BibitemShut {NoStop}%
\bibitem [{\citenamefont {Yost}\ \emph {et~al.}(2012)\citenamefont {Yost},
  \citenamefont {Hontz}, \citenamefont {Yeganeh},\ and\ \citenamefont
  {Voorhis}}]{Yost12}%
  \BibitemOpen
  \bibfield  {author} {\bibinfo {author} {\bibfnamefont {S.~R.}\ \bibnamefont
  {Yost}}, \bibinfo {author} {\bibfnamefont {E.}~\bibnamefont {Hontz}},
  \bibinfo {author} {\bibfnamefont {S.}~\bibnamefont {Yeganeh}}, \ and\
  \bibinfo {author} {\bibfnamefont {T.~V.}\ \bibnamefont {Voorhis}},\ }\href
  {\doibase 10.1021/jp304433t} {\bibfield  {journal} {\bibinfo  {journal} {The
  Journal of Physical Chemistry C}\ }\textbf {\bibinfo {volume} {116}},\
  \bibinfo {pages} {17369} (\bibinfo {year} {2012})}\BibitemShut {NoStop}%
\bibitem [{\citenamefont {Bossanyi}\ \emph {et~al.}(2021)\citenamefont
  {Bossanyi}, \citenamefont {Matthiesen}, \citenamefont {Wang}, \citenamefont
  {Smith}, \citenamefont {Kilbride}, \citenamefont {Shipp}, \citenamefont
  {Chekulaev}, \citenamefont {Holland}, \citenamefont {Anthony}, \citenamefont
  {Zaumseil} \emph {et~al.}}]{bossanyi2021}%
  \BibitemOpen
  \bibfield  {author} {\bibinfo {author} {\bibfnamefont {D.~G.}\ \bibnamefont
  {Bossanyi}}, \bibinfo {author} {\bibfnamefont {M.}~\bibnamefont
  {Matthiesen}}, \bibinfo {author} {\bibfnamefont {S.}~\bibnamefont {Wang}},
  \bibinfo {author} {\bibfnamefont {J.~A.}\ \bibnamefont {Smith}}, \bibinfo
  {author} {\bibfnamefont {R.~C.}\ \bibnamefont {Kilbride}}, \bibinfo {author}
  {\bibfnamefont {J.~D.}\ \bibnamefont {Shipp}}, \bibinfo {author}
  {\bibfnamefont {D.}~\bibnamefont {Chekulaev}}, \bibinfo {author}
  {\bibfnamefont {E.}~\bibnamefont {Holland}}, \bibinfo {author} {\bibfnamefont
  {J.~E.}\ \bibnamefont {Anthony}}, \bibinfo {author} {\bibfnamefont
  {J.}~\bibnamefont {Zaumseil}},  \emph {et~al.},\ }\href@noop {} {\bibfield
  {journal} {\bibinfo  {journal} {Nature chemistry}\ }\textbf {\bibinfo
  {volume} {13}},\ \bibinfo {pages} {163} (\bibinfo {year} {2021})}\BibitemShut
  {NoStop}%
\bibitem [{\citenamefont {Irkhin}\ \emph {et~al.}(2016)\citenamefont {Irkhin},
  \citenamefont {Biaggio}, \citenamefont {Zimmerling}, \citenamefont
  {D{\"o}beli},\ and\ \citenamefont {Batlogg}}]{Irkhin16}%
  \BibitemOpen
  \bibfield  {author} {\bibinfo {author} {\bibfnamefont {P.}~\bibnamefont
  {Irkhin}}, \bibinfo {author} {\bibfnamefont {I.}~\bibnamefont {Biaggio}},
  \bibinfo {author} {\bibfnamefont {T.}~\bibnamefont {Zimmerling}}, \bibinfo
  {author} {\bibfnamefont {M.}~\bibnamefont {D{\"o}beli}}, \ and\ \bibinfo
  {author} {\bibfnamefont {B.}~\bibnamefont {Batlogg}},\ }\href@noop {}
  {\bibfield  {journal} {\bibinfo  {journal} {Applied Physics Letters}\
  }\textbf {\bibinfo {volume} {108}},\ \bibinfo {pages} {063302} (\bibinfo
  {year} {2016})}\BibitemShut {NoStop}%
\bibitem [{\citenamefont {Seki}\ \emph {et~al.}(2021)\citenamefont {Seki},
  \citenamefont {Yoshida}, \citenamefont {Yago}, \citenamefont {Wakasa},\ and\
  \citenamefont {Katoh}}]{Seki21}%
  \BibitemOpen
  \bibfield  {author} {\bibinfo {author} {\bibfnamefont {K.}~\bibnamefont
  {Seki}}, \bibinfo {author} {\bibfnamefont {T.}~\bibnamefont {Yoshida}},
  \bibinfo {author} {\bibfnamefont {T.}~\bibnamefont {Yago}}, \bibinfo {author}
  {\bibfnamefont {M.}~\bibnamefont {Wakasa}}, \ and\ \bibinfo {author}
  {\bibfnamefont {R.}~\bibnamefont {Katoh}},\ }\bibfield  {booktitle} {\emph
  {\bibinfo {booktitle} {The Journal of Physical Chemistry C}},\ }\href
  {\doibase 10.1021/acs.jpcc.0c10582} {\bibfield  {journal} {\bibinfo
  {journal} {The Journal of Physical Chemistry C}\ }\textbf {\bibinfo {volume}
  {125}},\ \bibinfo {pages} {3295} (\bibinfo {year} {2021})}\BibitemShut
  {NoStop}%
\end{thebibliography}%

\end{document}